\documentclass[]{spie}  

\usepackage{amsmath,amsfonts,amssymb}
\usepackage{graphicx}
\usepackage{epstopdf}
\usepackage[colorlinks=true, allcolors=blue]{hyperref}
\usepackage{bm}

\title{Spin-transfer torque in ferromagnetic bilayers generated by anomalous Hall effect and anisotropic magnetoresistance}

\author[a]{Tomohiro Taniguchi}
\author[b]{Julie Grollier}
\author[c]{M. D. Stiles}
\affil[a]{National Institute of Advanced Industrial Science and Technology (AIST), Spintronics Research Center, Tsukuba, Ibaraki 305-8568, Japan}
\affil[b]{Unit\'e Mixte de Physique CNRS/Thales and Universit\'e Paris Sud 11, 1 Avenue Fresnel, 91767 Palaiseau, France}
\affil[c]{Center for Nanoscale Science and Technology, National Institute of Standards and Technology, Gaithersburg, Maryland 20899-6202, USA}

\begin{document}
\maketitle

\begin{abstract}
We propose an experimental scheme to determine the spin-transfer torque efficiency excited by the spin-orbit interaction
in ferromagnetic bilayers from the measurement of the longitudinal magnetoresistace.
Solving a diffusive spin-transport theory with appropriate boundary conditions gives
an analytical formula of the longitudinal charge current density.
The longitudinal charge current has a term that is proportional to the square of the spin-transfer torque efficiency and
that also depends on the ratio of the film thickness to the spin diffusion length of the ferromagnet.
Extracting this contribution from measurements of the longitudinal resistivity as a function of the thickness
can give the spin-transfer torque efficiency.
\end{abstract}

\keywords{Spintronics, spin-orbit torque, anomalous Hall effect, anisotropic magnetoresistance}


\section{INTRODUCTION}
\label{sec:introduction}  

Spin-dependent electron transport in nanostructured ferromagnetic/nonmagnetic multilayers shows
interesting phenomena such as magnetoresistance effects and spin-transfer torque excitation
\cite{baibich88,binasch89,camley89,barnas90,zhang91,pratt93,valet93,slonczewski96,berger96,katine00,grollier01,stiles02,kiselev03,krivorotov05,bass07}.
Many of these phenomena are well described by diffusive spin transport theory
together with with spin-dependent interface scattering theory \cite{brataas01,bauer03}.
Early investigations focused on longitudinal transport, i.e., the transport along the direction of an external electric field.
Recent efforts have focused on the transverse transport, particularly the spin Hall effect.
The spin Hall effect \cite{dyakonov71,hirsch99} in nonmagnetic heavy metals originates when the spin-orbit interaction scatters the spin-up and spin-down electrons in opposite directions,
generating pure spin currents in the transverse direction.
These spin currents excite spin-transfer torques in the attached ferromagnetic layers \cite{liu12,liu12a,garello13,haney13,haney13a,kim14,qiu14,cubukcu14,yu14}
or provide magnetoresistance effects in both longitudinal and transverse directions \cite{weiler12,hahn13,althammer13,nakayama13,chen13,cho15,kim16}.


The spin-orbit interaction in ferromagnets also generates spin currents through the anomalous Hall effect and the anisotropic magnetoresistance \cite{pugh53,mcguire75}.
While the spin polarizations of the spin currents in nonmagnets due to the spin Hall effect are geometrically fixed,
the spin polarizations in ferromagnets are set by the magnetization direction because the large exchange interaction in the ferromagnet results in dephasing of any transverse spins.
This dependence allows for manipulation of the spin polarization by controlling the direction of the magnetization.
Therefore, a wide variety of magnetization dynamics can be excited by spin-transfer torques generated by spin-orbit interactions in ferromagnets \cite{taniguchi15PRApplied}.
A key quantity in these magnetization dynamics is the efficiency of the spin-transfer torque,
which depends on the spin polarizations of the ferromagnet and the transverse conductivity.
In this paper, we propose an experimental scheme to determine this efficiency from measurements of the longitudinal magnetoresistance.


\section{SPIN TRANSPORT THEORY IN DIFFUSIVE FERROMAGNETS}
\label{sec:Spin Transport Theory in Diffusive Ferromagnets}

In this section, we give the definition of the charge and spin currents in ferromagnets that show
the anomalous Hall effect and anisotropic magnetoresistance.
Then, we use these to propose an experimental scheme to determine the spin-transfer torque efficiency due to the anomalous Hall effect
by measuring the longitudinal resistivity as a function of film thickness.


\subsection{Definition of spin-dependent current}

The charge and spin currents in ferromagnets are combinations of the current carried by the spin-up ($\uparrow$) and spin-down ($\downarrow$) electrons.
In the presence of the anomalous Hall effect and anisotropic magnetoresistance,
the current densities of the spin-$s$ ($s=\uparrow,\downarrow$) electrons are given by \cite{taniguchi15PRApplied}
\begin{equation}
\begin{split}
  \mathbf{j}^{\uparrow}
  =&
  \frac{(1+\beta) \sigma}{2e}
  \bm{\nabla}
  \mu_{\uparrow}
  +
  \frac{(1+\zeta) \sigma_{\rm AH}}{2e}
  \mathbf{m}
  \times
  \bm{\nabla}
  \mu_{\uparrow}
  +
  \frac{(1+\eta) \sigma_{\rm AMR}}{2e}
  \mathbf{m}
  \left(
    \mathbf{m}
    \cdot
    \bm{\nabla}
    \mu_{\uparrow}
  \right),
  \label{eq:current_up}
\end{split}
\end{equation}
\begin{equation}
\begin{split}
  \mathbf{j}^{\downarrow}
  =&
  \frac{(1-\beta) \sigma}{2e}
  \bm{\nabla}
  \mu_{\downarrow}
  +
  \frac{(1-\zeta) \sigma_{\rm AH}}{2e}
  \mathbf{m}
  \times
  \bm{\nabla}
  \mu_{\downarrow}
  +
  \frac{(1-\eta) \sigma_{\rm AMR}}{2e}
  \mathbf{m}
  \left(
    \mathbf{m}
    \cdot
    \bm{\nabla}
    \mu_{\downarrow}
  \right),
  \label{eq:current_down}
\end{split}
\end{equation}
where $\sigma$, $\sigma_{\rm AH}$, and $\sigma_{\rm AMR}$ are
the longitudinal conductivity and the conductivities due to the anomalous Hall effect and the anisotropic magnetoresistance, respectively,
and $\beta$, $\zeta$, and $\eta$ are the spin polarizations of the respective currents.
The electron charge is $-e$.
The electrochemical potential of the spin-$s$ electrons is denoted as $\mu_{s}$.
The unit vector pointing in the direction of the magnetization is $\mathbf{m}$.
The total electric current density $\mathbf{j}=\mathbf{j}^{\uparrow}+\mathbf{j}^{\downarrow}$ is
\begin{equation}
\begin{split}
  \mathbf{j}
  =&
  \frac{\sigma}{e}
  \bm{\nabla}
  \bar{\mu}
  +
  \frac{\beta \sigma}{e}
  \bm{\nabla}
  \delta
  \mu
  +
  \frac{\sigma_{\rm AH}}{e}
  \mathbf{m}
  \times
  \bm{\nabla}
  \bar{\mu}
  +
  \frac{\zeta \sigma_{\rm AH}}{e}
  \mathbf{m}
  \times
  \bm{\nabla}
  \delta
  \mu
  +
  \frac{\sigma_{\rm AMR}}{e}
  \bm{\nabla}
  \mathbf{m}
  \left(
    \mathbf{m}
    \cdot
    \bm{\nabla}
    \bar{\mu}
  \right)
  +
  \frac{\eta \sigma_{\rm AMR}}{e}
  \mathbf{m}
  \left(
    \mathbf{m}
    \cdot
    \bm{\nabla}
    \delta
    \mu
  \right),
  \label{eq:charge_current}
\end{split}
\end{equation}
where we define $\bar{\mu}=(\mu_{\uparrow}+\mu_{\downarrow})/2$ and $\delta\mu=(\mu_{\uparrow}-\mu_{\downarrow})/2$.
The tensor spin current density is the outer product
of the spin polarization and the current direction,
i.e., $\mathbf{Q}=-[\hbar/(2e)]\mathbf{m} \otimes (\mathbf{j}^{\uparrow}-\mathbf{j}^{\downarrow})$,
where $\mathbf{j}^{\uparrow}-\mathbf{j}^{\downarrow}$ is
\begin{equation}
\begin{split}
  \mathbf{j}^{\uparrow}
  -
  \mathbf{j}^{\downarrow}
  =&
  \frac{\sigma}{e}
  \bm{\nabla}
  \delta
  \mu
  +
  \frac{\beta \sigma}{e}
  \bm{\nabla}
  \bar{\mu}
  +
  \frac{\sigma_{\rm AH}}{e}
  \mathbf{m}
  \times
  \bm{\nabla}
  \delta
  \mu
  +
  \frac{\zeta \sigma_{\rm AH}}{e}
  \mathbf{m}
  \times
  \bm{\nabla}
  \bar{\mu}
  +
  \frac{\sigma_{\rm AMR}}{e}
  \bm{\nabla}
  \mathbf{m}
  \left(
    \mathbf{m}
    \cdot
    \bm{\nabla}
    \delta
    \mu
  \right)
  +
  \frac{\eta \sigma_{\rm AMR}}{e}
  \mathbf{m}
  \left(
    \mathbf{m}
    \cdot
    \bm{\nabla}
    \bar{\mu}
  \right).
\end{split}
\end{equation}
Here, we assume that the spin polarization of the spin current in the ferromagnet is parallel to the magnetization because the large exchange coupling between the spin current and the magnetization results in fast dephasing of any transverse spins
\cite{slonczewski96,stiles02,brataas01,zhang02,zhang04,taniguchi08,ghosh12}.


\subsection{Determination of spin-transfer torque efficiency by measuring longitudinal magnetoresistance}

In trilayer structures with two ferromagnetic layers separated by a non-magnetic spacer layer, which breaks the direct exchange between the ferromagnets, currents flowing in the plane of the structure generate spin currents perpendicular to the film plane through the the anomalous Hall effect and the anisotropic magnetoresistance.  Some of the spin currents generated in each layer cross the spacer layer, thereby injecting a spin current into the other layer 
and exciting spin-transfer torques.
Here, for simplicity, we focus on the anomalous Hall effect alone.
As discussed in Sec. III of our previous work\cite{taniguchi15PRApplied},
the magnitude of the spin-transfer torque, $\mathbf{T}$, on the magnetization of one layer excited by the anomalous Hall effect in the other layer is
\begin{equation}
  |\mathbf{T}|
  \propto
  \left(
    \beta
    -
    \zeta
  \right)
  \frac{\sigma_{\rm AH}}{\sigma}.
\end{equation}
Therefore, it is important to estimate this factor in real materials
for accurate prediction of spin-transfer torques.
For convenience, we refer to $(\beta-\zeta)(\sigma_{\rm AH}/\sigma)$ as the spin-transfer torque efficiency,
as it plays a role similar to the spin Hall angle in nonmagnets.

\begin{figure}
\centerline{\includegraphics[width=0.8\columnwidth]{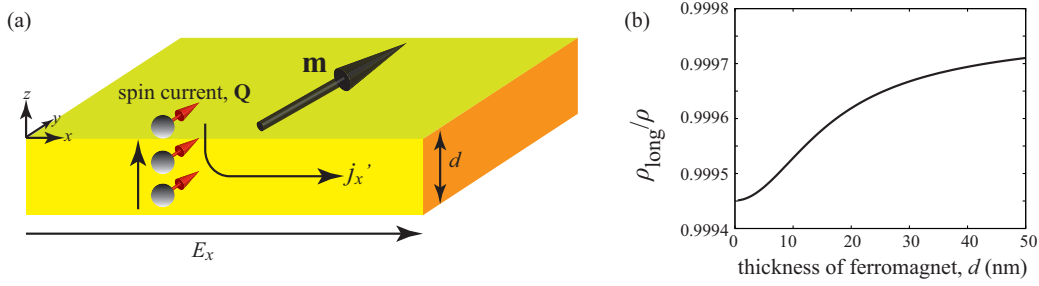}}
\caption{
        (a) A schematic view of the system under consideration.
            An external electric field $E_{x}$ generates a spin current $\mathbf{Q}$ flowing in the $z$-direction,
            and the spin current is converted to a charge current $j_{x}^{\prime}$.
        (b) The longitudinal resistivity $\rho_{\rm long}$ normalized by the resistivity $\rho$ as a function of the thickness of the ferromagnet $d$.
         }
\label{fig:fig1}
\end{figure}

In this section, we briefly discuss an experimental scheme to estimate the spin-transfer torque efficiency.
The basic idea of the present proposal is similar to the spin Hall magnetoresistance \cite{chen13},
where an applied charge current generates a spin current by the spin Hall effect,
and the spin current is converted to an additional electric current by the inverse spin Hall effect.
As a result, an extra longitudinal resistivity appears, which depends on the ratio between the thickness and the spin diffusion length of a nonmagnet.
Similarly, in a ferromagnetic layer, an extra resistivity due to the conversion between the charge and spin currents appears due to the anomalous Hall effect.
Figure \ref{fig:fig1}(a) schematically shows the system under consideration.
Let us consider the electron flow in a ferromagnet,
where an external electric field $E_{x}$ is applied along the $x$-direction,
and thus, $\partial_{x}(\bar{\mu}/e)=E_{x}$.
We assume translational symmetry along the $y$-direction,
and therefore, the spin accumulation spatially varies only along the $z$-direction.
The spin current $\mathbf{Q}$ is generated along the $z$-direction by the anomalous Hall effect,
which creates a spin accumulation $\delta\mu$.
This spin current is again converted to a charge current flowing in the $x$-direction, as discussed below.



To calculate the longitudinal resistivity in a ferromagnet, we start with
the charge current density along the $z$-direction obtained from Eq. (\ref{eq:charge_current}) given by
\begin{equation}
  j_{z}
  =
  \frac{\sigma}{e}
  \partial_{z}
  \bar{\mu}
  -
  \sigma_{\rm AH}
  E_{x}
  m_{y}
  +
  \frac{\beta \sigma}{e}
  \partial_{z}
  \delta
  \mu.
\end{equation}
We apply the open circuit boundary conditions along the $z$-direction, i.e., $j_{z}=0$
and $\partial_{z}(\bar{\mu}+\beta\delta\mu)=[(\sigma_{\rm AH}m_{y})/\sigma]eE_{x}$.
Then, we find that $\bar{\mu}$ and $\delta\mu$ are related as
\begin{equation}
  \bar{\mu}
  =
  -\beta
  \delta
  \mu
  +
  eE_{x}
  x
  +
  \frac{\sigma_{\rm AH} m_{y}}{\sigma}
  eE_{x}
  z.
\end{equation}
Using this equation, the charge current density along the $x$-direction obtained from Eq. (\ref{eq:charge_current}) becomes
\begin{equation}
  j_{x}
  =
  \sigma
  \left[
    1
    +
    \left(
      \frac{\sigma_{\rm AH} m_{y}}{\sigma}
    \right)^{2}
  \right]
  E_{x}
  -
  \frac{(\beta-\zeta)\sigma_{\rm AH} m_{y}}{e}
  \left(
    \partial_{z}
    \delta
    \mu
  \right).
  \label{eq:current_x}
\end{equation}
The term $\sigma E_{x}$ is the conventional longitudinal charge current density,
whereas $\sigma(\sigma_{\rm AH}m_{y}/\sigma)^{2}E_{x}$ originates from the internal electric field, $(\sigma_{\rm AH}m_{y}/\sigma)E_{x}$, due to the open circuit boundary conditions
and the anomalous Hall effect.
On the other hand, the last term proportional to $\partial_{z}\delta\mu$ arises from the conversion of the spin current,
and depends on the spin-transfer torque efficiency.
We emphasize that this last term depends on the thickness of the ferromagnet along the $z$-direction,
whereas the other terms in Eq. (\ref{eq:current_x}) are independent of the thickness.
Therefore, we expect that the spin-transfer torque efficiency can be estimated
from the dependence of the charge current density along the $x$-direction on the ferromagnetic thickness.
We confirm this idea by the following calculation.
To simplify the notation, we denote the last term of Eq. (\ref{eq:current_x}) as $j_{x}^{\prime}$,
\begin{equation}
  j_{x}^{\prime}
  =
  -\frac{(\beta-\zeta)\sigma_{\rm AH} m_{y}}{e}
  \left(
    \partial_{z}
    \delta
    \mu
  \right).
  \label{eq:current_x_prime}
\end{equation}

The spin accumulation $\delta\mu$ obeys the diffusion equation given by \cite{valet93}
\begin{equation}
  \frac{d^{2}\delta\mu}{dz^{2}}
  =
  \frac{\delta \mu}{\ell^{2}},
  \label{eq:diffusion_equation}
\end{equation}
where $\ell$ is the spin diffusion length.
Applying open circuit boundary conditions to the spin current along the $z$-direction,
the solution of Eq. (\ref{eq:diffusion_equation}) is
\begin{equation}
  \delta
  \mu
  =
  \frac{(\beta-\zeta) e \ell \sigma_{\rm AH} m_{y} E_{x}}{(1-\beta^{2}) \sigma \sinh(d/\ell)}
  \left[
    \cosh
    \left(
      \frac{z-d}{\ell}
    \right)
    -
    \cosh
    \left(
      \frac{z}{\ell}
    \right)
  \right],
  \label{eq:spin_accumulation}
\end{equation}
where $d$ is the thickness of the ferromagnet along the $z$-direction,
and we assume that the ferromagnet lies between $0 \le z \le d$.
Substituting Eq. (\ref{eq:spin_accumulation}) into Eq. (\ref{eq:current_x_prime}),
the averaged charge current density is given by
\begin{equation}
\begin{split}
  \langle j_{x}^{\prime} \rangle
  & \equiv
  \frac{1}{d}
  \int_{0}^{d}
  j_{x}^{\prime}
  dz
\\
  &=
  -\frac{(\beta-\zeta) \sigma_{\rm AH} m_{y}}{ed}
  \left[
    \delta
    \mu(z=d)
    -
    \delta
    \mu(z=0)
  \right]
\\
  &=
  \frac{2 [(\beta-\zeta) \sigma_{\rm AH} m_{y}]^{2}E_{x} \ell}{(1-\beta^{2})\sigma d}
  \tanh
  \left(
    \frac{d}{2 \ell}
  \right).
  \label{eq:averaged_current}
\end{split}
\end{equation}
Thus, the total current density given by Eq. (\ref{eq:current_x}) becomes
\begin{equation}
  j_{x}
  =
  \sigma
  \left\{
    1
    +
    \left(
      \frac{\sigma_{\rm AH}}{\sigma}
      m_{y}
    \right)^{2}
    +
    \frac{2 \ell}{(1-\beta^{2})d}
    \left[
      \frac{(\beta-\zeta) \sigma_{\rm AH}}{\sigma}
    \right]^{2}
    m_{y}^{2}
    \tanh
    \left(
      \frac{d}{2 \ell}
    \right)
  \right\}
  E_{x}.
\end{equation}
Defining the longitudinal resistivity $\rho_{\rm long}$ as
$\rho_{\rm long}=(j_{x}/E_{x})^{-1}$, we find that
\begin{equation}
  \frac{\rho_{\rm long}}{\rho}
  \simeq
  1
  -
  \left(
    \frac{\sigma_{\rm AH}}{\sigma}
    m_{y}
  \right)^{2}
  -
  \frac{2\ell}{(1-\beta^{2})d}
  \left[
    \frac{(\beta-\zeta) \sigma_{\rm AH}}{\sigma}
  \right]^{2}
  m_{y}^{2}
  \tanh
  \left(
    \frac{d}{2\ell}
  \right),
  \label{eq:resistivity}
\end{equation}
where $\rho=1/\sigma$.
Figure \ref{fig:fig1}(b) shows an example of Eq. (\ref{eq:resistivity}), where the values of the parameters are
$\sigma_{\rm AH}/\sigma=0.015$, $\beta=0.40$, $\zeta=1.5$, and $\ell=5$ nm \cite{taniguchi15PRApplied,moritz08},
and $m_{y}$ is set to be $|m_{y}|=1$.
As mentioned above, only the last term depends on the thickness of the ferromagnet $d$,
therefore, this term can be isolated by measuring the dependence of the longitudinal resistivity on the thickness.
We emphasize that this contribution is proportional to the square
of the spin-transfer torque efficiency $(\beta-\zeta)\sigma_{\rm AH}/\sigma$.
Thus, the measurement of the longitudinal resistivity can be used to determine the spin-transfer torque efficiency.

There is a large complication to this approach to measuring the spin torque efficiency in ferromagnets.  Eq.~(\ref{eq:resistivity}) gives three contributions: one that is independent of thickness and magnetization direction, one that is independent of thickness but dependent on the magnetization direction, and the contribution of interest, which depends on both thickness and magnetization direction.  Unfortunately, there is an additional contribution that depends on the thickness.  This additional contribution arises from effects that are not captured in the diffusive approach used here.

While the diffusive transport theory developed here works well for many aspects of spin transport, it does not capture all of the important physics particularly for in-plane transport in thin films.  One such effect, originally described by Fuchs \cite{fuchs38} and Sondheimer \cite{sondheimer52}, is the extra resistance in thin films due to boundary scattering.  Such resistance effects are captured by approaches like the Boltzmann equation \cite{penn99}.  Another effect, is the inability of diffusive transport theories ito describe current-in-plane giant magnetoresistance (CIP-GMR), which can also be treated with a Boltzmann equation approach \cite{camley89,barnas90}.  Both of these effects occur on length scales set by the mean free path rather than the spin diffusion length.  The diffusive approach used here neglects all such effects.  Thus, the boundary scattering described by Fuchs and Sondheimer gives rise to thickness dependences with a similar form as those described here, but varying with the mean free path rather than the the spin diffusion length. In ferromagnets the situation is further complicated but the fact that there are different mean free paths for majority and minority electrons.  The mean free paths of typical ferromagnets, such as Fe, Co, Ni, and their alloys, are on the order of 0.1 nm to 1 nm \cite{gurney93},
whereas the spin diffusion lengths of these ferromagnets are on the order of 1 nm to 10 nm \cite{bass07};
for example, the spin-dependent mean free paths of the majority and minority spins are
5.5 nm and 0.6 nm for Co, 1.5 nm and 2.1 nm for Fe, and 4.6 nm and 0.6 nm for Ni${}_{80}$Fe${}_{20}$, respectively,
whereas the spin diffusion lengths are 40 nm for Co, 8.5 nm for Fe, and 5.5 nm for Ni${}_{80}$Fe${}_{20}$, respectively.

In the absence of spin-orbit scattering localized to the interface, we expect the boundary scattering contributions to be independent of the magnetization direction, making it possible, in principle, to separate them from the contribution of interest.  In that case, determining the spin torque efficiency requires measuring the thickness dependence accurately enough to separate the magnetization independent contribution due to boundary scattering from the magnetoresitive contribution due to the anomalous Hall effect.
This procedure would be easiest in ferromagnets with a large separation between the spin diffusion length and the mean free paths or in  films with interfaces that give almost purely specular scattering.


\section{Conclusion}
\label{sec:Conclusion}

In conclusion, we propose an experimental scheme to determine the spin-transfer torque efficiency
excited by the spin-orbit interaction in ferromagnets.
We derived the analytical formula of the longitudinal charge current density
by solving the diffusion equation of the spin accumulation with appropriate boundary conditions,
and found that the longitudinal current has a term proportional to the square of the spin-transfer torque efficiency.
This term depends on the ratio of the thickness and the spin diffusion length of the ferromagnet.
Therefore, the spin-transfer torque efficiency can be evaluated by measuring the dependence
of the longitudinal resistivity on the ferromagnetic thickness provided these effects can be separated from thickness dependent boundary scattering contributions.





\begin{thebibliography}{10}

\bibitem{baibich88}
Baibich, M.~N., Broto, J.~M., Fert, A., Dau, N.~V., Petroff, F., Eitenne, P.,
  Creuzet, G., and an~J.~Chazelas, A.~F., ``Giant {Magnetoresistance} of
  (001){Fe}/(001){Cr} {Magnetic} {Supperlattices},'' {\em Phys. Rev.
  Lett.}~{\bf 61},  2472 (1988).

\bibitem{binasch89}
Binasch, G., Gr\"uberg, P., Saurenbach, F., and Zinn, W., ``Enhanced
  magnetoresistance in layered magnetic structures with antiferromagnetic
  interlayer exchange,'' {\em Phys. Rev. B}~{\bf 39},  4828 (1989).

\bibitem{camley89}
Camley, R.~E. and Barn\'as, J., ``Theory of {Giant} {Magnetoresistance}
  {Effects} in {Magnetic} {Layered} {Structures} with {Antiferromagnetic}
  {Coupling},'' {\em Phys. Rev. Lett.}~{\bf 63},  664 (1989).

\bibitem{barnas90}
Barn\'as, J., Fuss, A., Camley, R.~E., Gr\"unberg, P., and Zinn, W., ``Novel
  magnetoresistance effect in layered magnetic structures: {Theory} and
  experiment,'' {\em Phys. Rev. B}~{\bf 42},  8110 (1990).

\bibitem{zhang91}
Zhang, S. and Levy, P.~M., ``Conductivity perpendicular to the plane of
  multilayered structures,'' {\em J. Appl. Phys.}~{\bf 69},  4786 (1991).

\bibitem{pratt93}
Pratt, W.~P., Lee, S.-F., Slaughter, J.~M., Loloee, R., Schroeder, P.~A., and
  Bass, J., ``Perpendicular {Giant} {Magnetoresistances} of {Ag}/{Co}
  {Multilayers},'' {\em Phys. Rev. Lett.}~{\bf 66},  3060 (1993).

\bibitem{valet93}
Valet, T. and Fert, A., ``Theory of the perpendicular magnetoresistance in
  magnetic multilayers,'' {\em Phys. Rev. B}~{\bf 48},  7099 (1993).

\bibitem{slonczewski96}
Slonczewski, J.~C., ``Current-driven excitation of magnetic multilayers,'' {\em
  J. Magn. Magn. Mater.}~{\bf 159},  L1 (1996).

\bibitem{berger96}
Berger, L., ``Emission of spin waves by a magnetic multilayer traversed by a
  current,'' {\em Phys. Rev. B}~{\bf 54},  9353 (1996).

\bibitem{katine00}
Katine, J.~A., Albert, F.~J., Buhrman, R.~A., Myers, E.~B., and Ralph, D.~C.,
  ``Current-{Driven} {Magnetization} {Reversal} and {Spin}-{Wave} {Excitations}
  in {Co}/ {Cu}/{Co} {Pillars},'' {\em Phys. Rev. Lett.}~{\bf 84},  3149
  (2000).

\bibitem{grollier01}
Grollier, J., Cros, V., George, A. H. J.~M., Jaffr\"es, H., and Fert, A.,
  ``Spin-polarized current induced switching in {Co}/{Cu}/{Co} pillars,'' {\em
  Appl. Phys. Lett.}~{\bf 78},  3663 (2001).

\bibitem{stiles02}
Stiles, M.~D. and Zangwill, A., ``Anatomy of spin-transfer torque,'' {\em Phys.
  Rev. B}~{\bf 66},  014407 (2002).

\bibitem{kiselev03}
Kiselev, S.~I., Sankey, J.~C., Krivorotov, I.~N., Emley, N.~C., Schoelkopf,
  R.~J., Buhrman, R.~A., and Ralph, D.~C., ``Microwave oscillations of a
  nanomagnet driven by a spin-polarized current,'' {\em Nature}~{\bf 425},  380
  (2003).

\bibitem{krivorotov05}
Krivorotov, I.~N., Emley, N.~C., Sankey, J.~C., Kiselev, S.~I., Ralph, D.~C.,
  and Buhrman, R.~A., ``Time-{Domain} {Measurements} of {Nanomagnet} {Dynamics}
  {Driven} by {Spin}-{Transfer} {Torques},'' {\em Science}~{\bf 307},  228
  (2005).

\bibitem{bass07}
Bass, J. and W.~P.~Pratt, J., ``Spin-diffusion lengths in metals and alloys,
  and spin-flipping at metal/metal interface: an experimentalist's critical
  review,'' {\em J. Phys.: Condens. Matter}~{\bf 19},  183201 (2007).

\bibitem{brataas01}
Brataas, A., Nazarov, Y.~V., and Bauer, G. E.~W., ``Spin-transport in
  multi-terminal normal metal-ferromagnet systems with non-colliner
  magnetizations,'' {\em Eur. Phys. J. B}~{\bf 22},  99 (2001).

\bibitem{bauer03}
Bauer, G. E.~W., Tserkovnyak, Y., Hernando, D.~H., and Brataas, A., ``Universal
  angular magnetoresistance and spin torque in ferromagnetic/normal metal
  hybrids,'' {\em Phys. Rev. B}~{\bf 67},  094421 (2003).

\bibitem{dyakonov71}
Dyakonov, M.~I. and Perel, V.~I., ``Current-induced spin orientation of
  electrons in semiconductors,'' {\em Phys. Lett. A}~{\bf 35},  459 (1971).

\bibitem{hirsch99}
Hirsch, J.~E., ``Spin {Hall} {Effect},'' {\em Phys. Rev. Lett.}~{\bf 83},  1834
  (1999).

\bibitem{liu12}
Liu, L., Lee, O.~J., Gudmundsen, T.~J., Ralph, D.~C., and Buhrman, R.~A.,
  ``Current-{Induced} {Switching} of {Perpendicularly} {Magnetized} {Magnetic}
  {Layers} {Using} {Spin} {Torque} from the {Spin} {Hall} {Effect},'' {\em
  Phys. Rev. Lett.}~{\bf 109},  096602 (2012).

\bibitem{liu12a}
Liu, L., Pai, C.-F., Li, Y., Tseng, H.~W., Ralph, D.~C., and Buhrman, R.~A.,
  ``Spin-{Torque} {Switching} with the {Giant} {Spin} {Hall} {Effect} of
  {Tantalum},'' {\em Science}~{\bf 336},  555 (2012).

\bibitem{garello13}
Garello, K., Miron, I.~M., Avci, C.~O., Freimuth, F., Mokrousov, Y., Bl\"ugel,
  S., Auffret, S., Boulle, O., Gaudin, G., and Gambardella, P., ``Symmetry and
  magnitude of spin-orbit torques in ferromagnetic heterostructures,'' {\em
  Nat. Nanotech.}~{\bf 8},  587 (2013).

\bibitem{haney13}
Haney, P.~M., Lee, H.-W., Lee, K.-J., Manchon, A., and Stiles, M.~D., ``Current
  induced torques and interfacial spin-orbit coupling: {Semiclassical}
  modeling,'' {\em Phys. Rev. B}~{\bf 87},  174411 (2013).

\bibitem{haney13a}
Haney, P.~M., Lee, H.-W., Lee, K.-J., Manchon, A., and Stiles, M.~D.,
  ``Current-induced torques and interfacial spin-orbit coupling,'' {\em Phys.
  Rev. B}~{\bf 88},  214417 (2013).

\bibitem{kim14}
Kim, J., Sinha, J., Mitani, S., Hayashi, M., Takahashi, S., Maekawa, S.,
  Yamanouchi, M., and Ohno, H., ``Anomalous temperature dependence of
  current-induced torques in {Co}{Fe}{B}/{Mg}{O} heterostructures with
  {Ta}-based underlayers,'' {\em Phys. Rev. B}~{\bf 89},  174424 (2014).

\bibitem{qiu14}
Qiu, X., Deorani, P., Narayanapillai, K., Lee, K.-S., Lee, K.-J., Lee, H.-W.,
  and Yang, H., ``Angular and temperature dependence of current induced
  spin-orbit effective fields in {Ta}/{Co}{Fe}{B}/{Mg}{O} nanowires,'' {\em
  Sci. Rep.}~{\bf 4},  4491 (2014).

\bibitem{cubukcu14}
Cubukcu, M., Boulle, O., Drouard, M., Garello, K., Avci, C.~O., Miron, I.~M.,
  Langer, J., Ocker, B., Gambardella, P., and Gaudin, G., ``Spin-orbit torque
  magnetization switching of a three-terminal perpendicular magnetic tunnel
  junction,'' {\em Appl. Phys. Lett.}~{\bf 104},  042406 (2014).

\bibitem{yu14}
Yu, G., Upadhyaya, P., Fan, Y., Alzate, J., Jiang, W., Wong, K.~L., Takei, S.,
  Bender, S.~A., Chang, L.-T., Jiang, Y., Lang, M., Tang, J., Wang, Y.,
  Tserkovnyak, Y., Amiri, P.~K., and Wang, K.~L., ``Switching of perpendicular
  magnetization by spin-orbit torques in the absence of external magnetic
  fields,'' {\em Nat. Nanotech.}~{\bf 9},  548 (2014).

\bibitem{weiler12}
Weiler, M., Althammer, M., Czeschka, F.~D., Huebl, H., Wagner, M.~S., Opel, M.,
  Imort, I.-M., Reiss, G., Thomas, A., Gross, R., and Goennewein, S. T.~B.,
  ``Local {Charge} and {Spin} {Currents} in {Magnetothermal} {Landscape},''
  {\em Phys. Rev. Lett.}~{\bf 108},  106602 (2012).

\bibitem{hahn13}
Hahn, C., de~Loubens, G., Klein, O., Viret, M., Naletov, V.~V., and BenYoussef,
  J., ``Comparative measurements of inverse spin {Hall} effects and
  magnetoresistance in {Y}{I}{G}/{Pt} and {Y}{I}{G}/{Ta},'' {\em Phys. Rev.
  B}~{\bf 87},  174417 (2013).

\bibitem{althammer13}
Althammer, M., Meyer, S., Nakayama, H., Schreier, M., Altmannshofer, S.,
  Weiler, M., Huebl, H., Gepr\"ags, S., Opel, M., Gross, R., Meier, D., Klewe,
  C., Kuschel, T., Schmalhorst, J.-M., Reiss, G., Shen, L., Gupta, A., Chen,
  Y.-T., Bauer, G. E.~W., Saitoh, E., and Goennenwein, S. T.~B., ``Quantitative
  study of the spin {Hall} magnetoresistance in ferromagnetic insulator/normal
  metal hybrids,'' {\em Phys. Rev. B}~{\bf 87},  224401 (2013).

\bibitem{nakayama13}
Nakayama, H., Althammer, M., Chen, Y.-T., Uchida, K., Kajiwara, Y., Kikuchi,
  D., Ohtani, T., Gepr\"ags, S., Opel, M., Takahashi, S., Gross, R., Bauer, G.
  E.~W., Goennenwein, S. T.~B., and Saitoh, E., ``Spin {Hall}
  {Magnetoresistance} {Induced} by a {Nonequilibrium} {Proximity} {Effect},''
  {\em Phys. Rev. Lett.}~{\bf 110},  206601 (2013).

\bibitem{chen13}
Chen, W.~T., Takahashi, S., Nakayama, H., Althammer, M., Goennewein, S. T.~B.,
  Saitoh, E., and Bauer, G. E.~W., ``Theory of spin {Hall} magntoresistance,''
  {\em Phys. Rev. B}~{\bf 87},  144411 (2013).

\bibitem{cho15}
Cho, S., Baek, S.-H.~C., Lee, K.-D., Jo, Y., and Park, B.-G., ``Large spin
  {Hall} magnetoresistance and its correlation to the spin-orbit torque in
  {W}/{Co}{Fe}{B}/{Mg}{O} structures,'' {\em Sci. Rep.}~{\bf 5},  14668 (2015).

\bibitem{kim16}
Kim, J., Sheng, P., Takahashi, S., Mitani, S., and Hayashi, M., ``Spin {Hall}
  {Magnetoresistance} in {Metallic} {Bilayers},'' {\em Phys. Rev. Lett.}~{\bf
  116},  097201 (2016).

\bibitem{pugh53}
Pugh, E.~M. and Postoker, N., ``Hall {Effect} in {Ferromagnetic} {Materials},''
  {\em Rev. Mod. Phys.}~{\bf 25},  151 (1953).

\bibitem{mcguire75}
McGuire, T.~R. and Potter, R., ``Anisotropic {Magnetoresistance} in
  {Ferromagnetic} $3d$ {Alloys},'' {\em IEEE Trans. Magn.}~{\bf 11},  1018
  (1975).

\bibitem{taniguchi15PRApplied}
Taniguchi, T., Grollier, J., and Stiles, M.~D., ``Spin-{Transfer} {Torque}
  {Generated} by the {Anomalous} {Hall} {Effect} and {Anisotropic}
  {Magnetoresistance},'' {\em Phys. Rev. Applied}~{\bf 3},  044001 (2015).

\bibitem{zhang02}
Zhang, S., Levy, P.~M., and Fert, A., ``Mechanisms of {Spin}-{Polarized}
  {Current}-{Driven} {Magnetization} {Switching},'' {\em Phys. Rev. Lett.}~{\bf
  88},  236601 (2002).

\bibitem{zhang04}
Zhang, J., Levy, P.~M., Zhang, S., and Antropov, V., ``Identification of
  {Transverse} {Spin} {Currents} in {Noncollinear} {Magnetic} {Structures},''
  {\em Phys. Rev. Lett.}~{\bf 93},  256602 (2004).

\bibitem{taniguchi08}
Taniguchi, T., Yakata, S., Imamura, H., and Ando, Y., ``Determination of
  {Penetration} {Depth} of {Transverse} {Spin} {Current} in {Ferromagnetic}
  {Metals} by {Spin} {Pumping},'' {\em Appl. Phys. Express}~{\bf 1},  031302
  (2008).

\bibitem{ghosh12}
Ghosh, A., Auffret, S., Ebels, U., and Bailey, W.~E., ``Penetration {Depth} of
  {Transverse} {Spin} {Current} in {Ultrahin} {Ferromagnets},'' {\em Phys. Rev.
  Lett.}~{\bf 109},  127202 (2012).

\bibitem{moritz08}
Moritz, J., Rodmacq, B., Auffret, S., and Dieny, B., ``Extraordinary {Hall}
  effect in thin magnetic films and its potential for sensors, memories and
  magnetic logic applications,'' {\em J. Phys. D}~{\bf 41},  135001 (2008).

\bibitem{fuchs38}
Fuchs, K., ``The conductivity of thin metallic films according to the electron
  theory of metals,'' {\em Proc. Cambridge Philos. Soc.}~{\bf 34},  100 (1938).

\bibitem{sondheimer52}
Sondheimer, E.~H., ``The mean free path of electrons in metals,'' {\em Adv.
  Phys.}~{\bf 1},  1 (1952).

\bibitem{penn99}
Penn, D.~R. and Stiles, M.~D., ``Solution of the {Boltzmann} equation without
  the relaxation-time approximation,'' {\em Phys. Rev. B}~{\bf 59},  13338
  (1999).

\bibitem{gurney93}
Gurney, B.~A., Speriosu, V.~S., Nozieres, J.-P., Lefakis, H., Wilhoit, D.~R.,
  and Need, O.~U., ``Direct {Measurement} of {Spin}-{Dependent}
  {Conduction}-{Electron} {Mean} {Free} {Paths} in {Ferromagnetic} {Metals},''
  {\em Phys. Rev. Lett.}~{\bf 93},  4023 (1993).

\end{thebibliography}


\end{document}